\documentclass{article}
\usepackage{spconf,amsmath,graphicx,hyperref}
\usepackage{amssymb}
\usepackage{xcolor, xspace}         
\usepackage{multicol}
\usepackage{algorithm,algorithmicx,algpseudocode}
\usepackage{caption}
\usepackage{tikz}
\usepackage{pgfplots}
\pgfplotsset{compat=1.18}
\usepackage{subcaption} 
\usepackage{booktabs}
\usepackage{enumitem}

\global\definecolor{iorange}{RGB}{255,149,0}
\global\definecolor{ired}{RGB}{255,59,48}
\global\definecolor{igreen}{RGB}{52,199,89}
\global\definecolor{icyan}{RGB}{50,173,230}
\global\definecolor{iblue}{RGB}{0,122,255}
\global\definecolor{iyellow}{RGB}{255,204,10}
\global\definecolor{ipurple}{RGB}{175,82,222}
\def\etal{\textit{et al.}\xspace}

\title{Super Monotonic Alignment Search}
\name{Junhyeok Lee$^{1}$\sthanks{Work done at Supertone Inc.}, Hyeongju Kim$^{2}$}
\address{$^{1}$ Center for Language and Speech Processing, 
    Johns Hopkins University, Baltimore, MD, USA\\$^{2}$ Supertone Inc., Seoul, Republic of Korea}

\begin{document}

\ninept
\maketitle
\begin{abstract}
Monotonic alignment search (MAS), introduced by Glow-TTS, is one of the most popular algorithm in text-to-speech to estimate unknown alignments between text and speech. Since this algorithm needs to search for the most probable alignment with dynamic programming by caching all possible paths, the time complexity of the algorithm is $O(T \times S)$, where $T$ is the length of text and $S$ is the length of speech representation. The authors of Glow-TTS run this algorithm on CPU, and while they mentioned it is difficult to parallelize, we found that MAS can be parallelized in text length dimension and CPU execution consumes an inordinate amount of time for inter-device copy. Therefore, we implemented a Triton kernel and PyTorch JIT script to accelerate MAS on GPU without inter-device copy. As a result, Super-MAS Triton kernel is up to 72 times faster in the extreme-length case. The code is available at \url{https://github.com/supertone-inc/super-monotonic-align}.
\end{abstract}


\section{Introduction}

GPU optimization plays a pivotal role in advancing modern speech processing and text-to-speech (TTS) systems, where large-scale models and real-time requirements demand considerable computational power and memory efficiency. 
Popular frameworks such as PyTorch \cite{pytorch} facilitate rapid prototyping, and emerging tools like \texttt{torch.compile} further accelerate workflows by capturing model graphs for optimized execution.
To further increase efficiency, developers leverage CUDA \cite{cuda} for low-level parallelization, as well as Triton-Lang \cite{triton} to write custom GPU kernels in a more flexible manner. 
Techniques like FlashAttention \cite{flashattn,flashattn2} exemplify specialized kernel optimizations that reduce memory overhead and accelerate attention modules in Transformer-based architectures \cite{transformer}, a common backbone for speech recognition and TTS tasks. 
Despite these advances, the inherent variability in the lengths of speech and text sequences during TTS training poses unique challenges for dynamic-length data. 
In particular, static graph optimization via compilation typically manages dynamic-length data by padding sequences to fixed sizes, which is commonly powers of two, to reduce the number of total required compilations. 
However, this approach is not suitable for TTS, where variability along two dimensions can lead to a number of compilations that exceeds the available cache size or force inefficient heavy zero paddings.

Monotonic alignment search (MAS) is an algorithm introduced by Kim \textit{et al.} \cite{glowtts}, which is applied to estimate unknown alignments between text and speech in a self-supervised manner.
Since this algorithm only needs text and speech pairs, many non-autoregressive TTS models are utilizing MAS during training \cite{glowtts,vits, pits, dualspeech}.
While original MAS utilized mel spectrogram, there are other studies \cite{pits, dualspeech} utilizing Yingram \cite{nansy} or NANSY feature \cite{nansypp} to MAS, we use the term speech representation instead of mel spectrogram.
Since this algorithm needs to search for the most probable alignment with dynamic programming by caching all paths, the time complexity of the algorithm is $O(T \times S)$, where $T$ is the length of text and $S$ is the length of speech representation. 
Official implementation of MAS\footnote{https://github.com/jaywalnut310/glow-tts} is implemented with Cython \cite{cython} and calculated on CPU with nested loops, while the authors mentioned it is difficult to parallelize.
However, we found that MAS can be parallelized in text length dimension and CPU execution is needed to copy large-size tensors between CPU and GPU, which consumes an inordinate amount of time.
Furthermore, since MAS is computed at every training iteration, it can become a significant bottleneck when training with long text and speech inputs.
Therefore, we implemented MAS with Triton kernel \cite{triton} and PyTorch JIT script \cite{pytorch} to accelerate MAS on GPU without the nested loops and inter-device copy.
Finally, Super-MAS Triton kernel is at least 19 times faster and up to 72 times faster than the original Cython implementation.

\begin{figure*}[ht]
    \centering
    \includegraphics[width=0.99\textwidth]{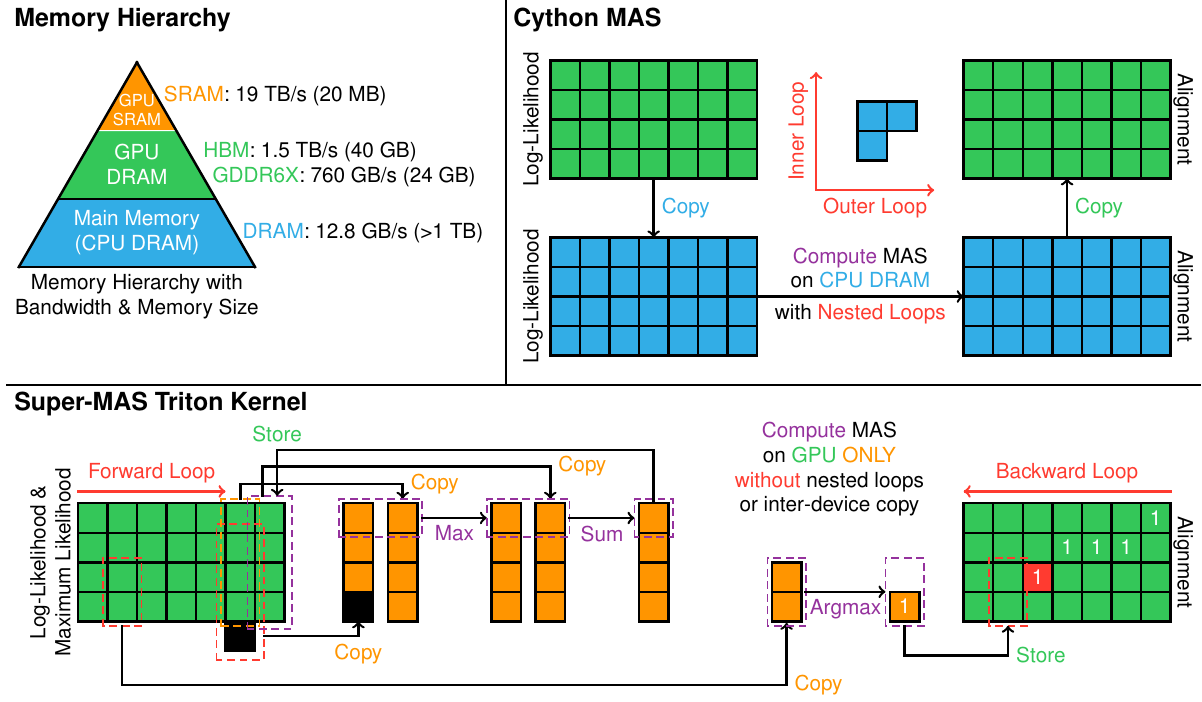}
    \caption{\textbf{Top Left}: Memory hierarchy with bandwidth \& memory size of Ampere architecture-based GPUs \cite{nvidia_a100,nvidia_ampere, flashattn}. 
    \textbf{Top Right}: Cython implementation of MAS including nested loops and inter-device copy. 
\textbf{Bottom}: Triton kernel implementation of MAS without nested loops or inter-device copy. The x-axis represents the speech length domain, while the y-axis represents the text length domain. Black block refers to the maximum negative value. The batch domain is not included for simplicity since both implementations run it in parallel.}    
\label{fig:memory}
\end{figure*}

\section{Monotonic Alignment Search} 
The original MAS calculation employs CPU executed nested loops as shown in Algorithm \ref{alg_mas}. According to the Kim \etal \cite{glowtts}, the procedure has a time complexity of $O(T\times S)$. Although the algorithm is not well suited to parallelization, it remains fast enough to run entirely on the CPU, requiring less than 20 ms per iteration—under 2 \% of the total training time. Moreover, MAS is only needed during training because the duration predictor provides the alignment at inference.

Despite these observations, our investigation uncovered several issues with the existing MAS implementation that warrant further discussion:
\begin{enumerate}[leftmargin=*]
    \item MAS can be parallelized in the text length dimension, while the original implementation uses nested loops.
    \item CPU execution consumes an inordinate amount of time for large inputs due to the need to copy large tensors between CPU and GPU.
    \item The time proportion spent on MAS increases as the length of the input text–speech pairs grows.
    \item The hard-coded value of \texttt{max\_neg\_val} at -1e9 is insufficient to prevent alignment mismatches.
\end{enumerate}

As mentioned in the introduction, the nested loops in the original MAS can be eliminated by parallelizing along the text length dimension, which corresponds to the inner loop. 
Algorithm \ref{alg_pmas} presents a parallelized version of MAS.
In the original implementation, the forward loop iterates over $j$ from 2 to $S$ (outer loop) and $i$ from 2 to $T$ (inner loop).
Since the iterations in the inner loop are independent, they can be vectorization through text length dimension for the parallel execution. 
With this parallelization, contrary to the statement from Glow-TTS, GPU computation becomes more efficient than CPU computation.
For brevity, $q_{i,j}$ denotes the log-likelihood matrix rather than $\log\mathcal{N}(z_j;\mu_i, \sigma_i)$.
In addition, inspired by FlashAttention \cite{flashattn}, we perform the computation inplace by using the same memory location in $q$, thereby preventing the allocation of new memory for the large intermediate tensors $Q$.
We implement this parallelized MAS in two frameworks to calculate on GPU, Triton-Lang \cite{triton} and PyTorch JIT script \cite{pytorch}.
Similar to the original implementation, the output alignment sizes of all implementations remain identical to the input log-likelihood sizes.

Although the MAS algorithm specifies masking certain values to negative infinity, the original implementation uses a hard-coded value of \texttt{max\_neg\_val} set to $-1\times 10^9$. 
While this could be replaced by infinity using \texttt{-torch.inf} in PyTorch or Triton, practical difficulties arise in Cython. 
Consequently, we opted to use a larger value for \texttt{max\_neg\_val}, specifically $-1\times 10^{32}$, for all implementations.


\subsection{Triton Kernel}
Figure \ref{fig:memory} provides a detailed illustration of how the Triton kernel calculates MAS.
In contrast to the Cython implementation, which parallelizes the batch dimension using a parallel for loop \texttt{prange}, Triton parallelizes the computation via a batch dimension–wise grid.
Because Triton requires the calculation block size to be a power of 2, we set it to the smallest power of 2 greater than $T$ and mask any excluded region using the constant \texttt{max\_neg\_val}.
The described algorithm is accelerated through Triton's just-in-time (JIT) compilation, which optimizes the execution at runtime.
We named this optimized kernel \textbf{Super-MAS}.

While the Triton kernel loads the block from GPU DRAM (HBM or GDDR) to GPU SRAM (L1 or L2 cache), it also has the capability to mask a region by assigning a specific value.
Therefore, unlike the Cython implementation, which starts with complex initialization by computing the cumulative sum for the first row and column, the Triton kernel only requires initializing the first column, thereby simplifying the initialization process.

\begin{algorithm}[t]
  \caption{Original Monotonic Alignment Search}
  \label{alg_mas}
\begin{algorithmic}
  \State {\bfseries Input:} log-likelihood matrix $q_{1:T,1:S}$, the speech representation length $S$, the text length $T$
  \State {\bfseries Output:} monotonic alignment $A^*$\\
  \State \textcolor{ired}{Initialize $Q_{1:T, 1:S} \leftarrow -\infty$, a cache to store the maximum log-likelihood calculations}
  \State \textcolor{ired}{Compute the first row $Q_{1, j} \leftarrow \sum_{k=1}^{j}{q_{i,k}}$, for all $j$}
  \For{$j=2$ {\bfseries to} $S$}
  \textcolor{ired}{
  \For{$i=2$ {\bfseries to} $\min(j,T)$}
  \State $Q_{i,j} \leftarrow \max(Q_{i -  1,j -  1}, Q_{i,j - 1}) + q_{i, j}$
  \EndFor}
  \EndFor
  \State Initialize $A^*(S) \leftarrow T$
  \For{$j=S-1$ {\bfseries to} $1$}
  \State $A^*(j) \leftarrow \mathrm{argmax}_{i \in \{A^*(j+1)-1, A^*(j+1)\}}Q_{i,j}$
  \EndFor
\end{algorithmic}
\end{algorithm}

A forward loop iterates over each column $j$ of the log-likelihood matrix, starting from the second column.
In each iteration, the function loads log-likelihood block from the current text position, denoted as $q_{1:T,j-1}$, as well as from the previous text position, represented by $q_{0:T-1, j-1}$, effectively processing block of data concurrently.
These blocks are then compared in element-wise, and the current block $q_{1:T, j}$ is updated inplace by adding the maximum of the two values to the corresponding element in the log-likelihood matrix, thereby computing the maximum likelihood for reaching that positions.
By processing data in blocks, the algorithm eliminates the inner loop and achieves the same computation more efficiently by parallelization.

Instead of original implementation allocates cumulative log-likelihood matrix separately to $Q_{1:T,1:S}$, we update it directly to $q_{1:T,1:S}$ to reduce overhead caused by additional memory allocation.
Similarly, to reduce additional memory allocation for alignment, we initially implemented a kernel to 
store a path matrix $A^*$ directly to $q$, we discovered that this approach is slow especially due to overwriting all values with zeros in certain regions.
This kernel resulted in significantly slower performance compared to our optimized implementation, so we held additional memory allocation for $A^*$ at the beginning and initialize it with zeros.

A backward loop iterates over each column $j$ to reconstruct the maximum path $A^*$.
To update only the alignment $A*(j) \leftarrow \mathrm{argmax}_{i \in \{A^*(j+1)-1, A^*(j+1)\}}q_{i,j}$, we initialize the maximum path matrix $A^*(j) \in \mathbb{R}^{T,S}$ as a full zero matrix.
The function then updates the path matrix by marking the elements along the maximum path based on comparisons between the left and those diagonally left-down in the value matrix.

The algorithm efficiently computes the maximum path by using parallelization to improve computational speed, leveraging the GPU's capabilities. 
Since the Triton kernel reuses the memory space of the given log-likelihood matrix by in-place computation, no additional memory is allocated and no memory overhead occurs. 
Furthermore, unlike the original Cython implementation, Super-MAS does not involve any inter-device communication between CPU and GPU memory.

\begin{table*}[ht]
\centering
\caption{
MAS execution times (ms). The fastest execution time is highlighted in bold, while the second fastest is underlined.}
  \resizebox{0.99 \linewidth}{!}{%
\begin{tabular}{c|*{16}{c}}
\toprule
\textbf{} & \textbf{128} & \textbf{256} & \textbf{384} & \textbf{512} & \textbf{640} & \textbf{768} & \textbf{896} & \textbf{1024} & \textbf{1152} & \textbf{1280} & \textbf{1408} & \textbf{1536} & \textbf{1664} & \textbf{1792} & \textbf{1920} & \textbf{2048} \\
\midrule
\textbf{Super-MAS} & \phantom{0}\textbf{0.4475} & \phantom{00}\textbf{1.617} & \phantom{00}\textbf{3.430} & \phantom{00}\textbf{5.839} & \phantom{00}\textbf{9.071} &  \phantom{0}\textbf{12.25} & \phantom{0}\textbf{15.20} & \phantom{00}\textbf{19.78} & \phantom{00}\textbf{33.28} & \phantom{00}\textbf{39.80} & \phantom{00}\textbf{47.46} & \phantom{00}\textbf{59.24} & \phantom{00}\textbf{70.07} & \phantom{00}\textbf{82.21} & \phantom{00}\textbf{99.63} & \phantom{0}\textbf{107.2}\\
\textbf{JIT\_v1} & 83.74\phantom{00} & 155.4\phantom{00} & 325.3\phantom{00} & 440.0\phantom{00} & 532.9\phantom{00} & 656.0\phantom{0} & \underline{558.0}\phantom{0} & \phantom{0}\underline{628.0}\phantom{0} & \phantom{0}\underline{706.0}\phantom{0} & \phantom{0}\underline{792.9}\phantom{0} & \phantom{0}\underline{903.8}\phantom{0} & \phantom{0}\underline{953.9}\phantom{0} & \underline{1031.}\phantom{00} & \underline{1558.}\phantom{00} & \underline{1183.}\phantom{00} & \underline{1262.}\phantom{0} \\
\textbf{JIT\_v2} & 53.22\phantom{00} & 104.6\phantom{00} & 237.8\phantom{00} & 344.7\phantom{00} & \underline{452.1}\phantom{00} & 587.2\phantom{0} & 620.1\phantom{0} & \phantom{0}815.9\phantom{0} & \phantom{0}968.5\phantom{0} & 1215.\phantom{00} & 1290.\phantom{00} & 1524.\phantom{00} & 2004.\phantom{00} & 2359.\phantom{00} & 2512.\phantom{00} & 2890.\phantom{0} \\
\textbf{Cython}  & \phantom{0}\underline{8.819}\phantom{0}  &  \phantom{0}\underline{43.53}\phantom{0}  & \underline{136.3}\phantom{00} & \underline{305.0}\phantom{00}  & 462.4\phantom{00} & \underline{488.2}\phantom{0}  & 863.9\phantom{0}  & 1299.\phantom{00}  & 1467.\phantom{00}  & 1930.\phantom{00}  & 2232.\phantom{00}  & 2959.\phantom{00}  & 3074.\phantom{00}  & 3931.\phantom{00}  & 4374.\phantom{00} & 7793.\phantom{0} \\
\bottomrule
\end{tabular}
}
\label{tab:result}
\end{table*}
    
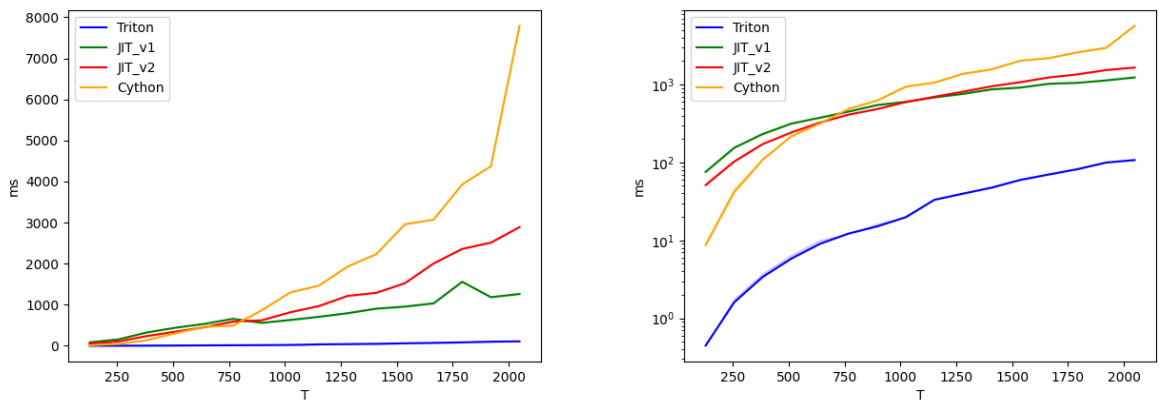
\begin{figure*}[ht]
    \centering
    \begin{subfigure}{0.49\textwidth}
    \centering
    \begin{tikzpicture}
        \begin{axis}[
            width=
            \linewidth,
            height=0.8 \linewidth,
            xlabel={T (Sequence Length)},
            ylabel={Execution Time (ms)},
            title={MAS Execution Time Comparison},
            legend pos=north west,
            grid=major,
            ymin=0,
            ymax=8000,
            xmode=linear, 
            ymode=linear, 
            log basis x=2,
            log basis y=10,
            xtick={256, 512, 768, 1024, 1280, 1536, 1792, 2048},
            xticklabels={256, 512, 768, 1024, 1280, 1536, 1792, 2048},
            legend style={draw=none, fill opacity=1.0, text opacity=1.0, fill=white, at={(0.02,0.98)}, anchor=north west,nodes={scale=0.7, transform shape}}
        ]

        \addplot[
            mark=o,
            color=iblue,
            thick
        ] coordinates {
            (128,0.447488) (256,1.616896) (384,3.430400) (512,5.838848) 
            (640,9.070592) (768,12.249088) (896,15.203328) (1024,19.778561) 
            (1152,33.276928) (1280,39.800835) (1408,47.456257) (1536,59.238914) 
            (1664,70.068741) (1792,82.205696) (1920,99.634689) (2048,107.218948)
        };
        \addlegendentry{Super-MAS}

        \addplot[
            mark=square,
            color=igreen,
            thick
        ] coordinates {
            (128,83.742203) (256,155.424774) (384,325.307404) (512,439.984131) 
            (640,532.910095) (768,655.960083) (896,557.997070) (1024,627.986450) 
            (1152,706.022400) (1280,792.861694) (1408,903.750671) (1536,953.907227) 
            (1664,1031.818237) (1792,1558.200317) (1920,1183.214600) (2048,1261.682739)
        };
        \addlegendentry{JIT\_v1}

        \addplot[
            mark=triangle,
            color=ired,
            thick
        ] coordinates {
            (128,53.222176) (256,104.632477) (384,237.820435) (512,344.654236) 
            (640,452.141907) (768,587.169739) (896,620.148315) (1024,815.933167) 
            (1152,968.533813) (1280,1215.021240) (1408,1289.656250) (1536,1523.870972) 
            (1664,2004.299438) (1792,2359.347900) (1920,2512.063477) (2048,2889.841797)
        };
        \addlegendentry{JIT\_v2}

        \addplot[
            mark=diamond,
            color=iyellow,
            thick
        ] coordinates {
            (128,8.819136) (256,43.533665) (384,136.257538) (512,304.981201) 
            (640,462.405304) (768,488.272858) (896,863.919067) (1024,1299.567871) 
            (1152,1467.056885) (1280,1930.171509) (1408,2231.598145) (1536,2959.377930) 
            (1664,3073.532471) (1792,3930.776367) (1920,4374.311035) (2048,7792.640137)
        };
        \addlegendentry{Cython}

        \end{axis}
    \end{tikzpicture}
\end{subfigure}\hfill
\begin{subfigure}{0.49\textwidth}
    \centering
    \begin{tikzpicture}
        \begin{axis}[
            width= \linewidth,
            height=0.8 \linewidth,
            xlabel={T (Sequence Length)},
            ylabel={Execution Time (ms)},
            title={MAS Execution Time Comparison},
            legend pos=south east,
            grid=major,
            ymin=0.1,
            ymax=10000,
            xmode=log, 
            ymode=log, 
            log basis x=2,
            log basis y=10,
            xtick={128, 256, 512, 1024, 2048},
            xticklabels={128, 256, 512, 1024, 2048},
            legend style={draw=none, fill opacity=1.0, text opacity=1.0, fill=white, at={(0.978,0.02)}, anchor=south east, nodes={scale=0.7, transform shape}}
        ]

        \addplot[
            mark=o,
            color=iblue,
            thick
        ] coordinates {
            (128,0.447488) (256,1.616896) (384,3.430400) (512,5.838848) 
            (640,9.070592) (768,12.249088) (896,15.203328) (1024,19.778561) 
            (1152,33.276928) (1280,39.800835) (1408,47.456257) (1536,59.238914) 
            (1664,70.068741) (1792,82.205696) (1920,99.634689) (2048,107.218948)
        };
        \addlegendentry{Super-MAS}

        \addplot[
            mark=square,
            color=igreen,
            thick
        ] coordinates {
            (128,83.742203) (256,155.424774) (384,325.307404) (512,439.984131) 
            (640,532.910095) (768,655.960083) (896,557.997070) (1024,627.986450) 
            (1152,706.022400) (1280,792.861694) (1408,903.750671) (1536,953.907227) 
            (1664,1031.818237) (1792,1558.200317) (1920,1183.214600) (2048,1261.682739)
        };
        \addlegendentry{JIT\_v1}

        \addplot[
            mark=triangle,
            color=ired,
            thick
        ] coordinates {
            (128,53.222176) (256,104.632477) (384,237.820435) (512,344.654236) 
            (640,452.141907) (768,587.169739) (896,620.148315) (1024,815.933167) 
            (1152,968.533813) (1280,1215.021240) (1408,1289.656250) (1536,1523.870972) 
            (1664,2004.299438) (1792,2359.347900) (1920,2512.063477) (2048,2889.841797)
        };
        \addlegendentry{JIT\_v2}

        \addplot[
            mark=diamond,
            color=iyellow,
            thick
        ] coordinates {
            (128,8.819136) (256,43.533665) (384,136.257538) (512,304.981201) 
            (640,462.405304) (768,488.272858) (896,863.919067) (1024,1299.567871) 
            (1152,1467.056885) (1280,1930.171509) (1408,2231.598145) (1536,2959.377930) 
            (1664,3073.532471) (1792,3930.776367) (1920,4374.311035) (2048,7792.640137)
        };
        \addlegendentry{Cython}
        \end{axis}
    \end{tikzpicture}
    \end{subfigure}
    \caption{MAS benchmark results in linear scale (left) and log scale (right). The batch size is fixed as 32, and the speech length is set to four times of text length.}
        \label{fig:result}
\end{figure*}

\subsection{PyTorch JIT Script}
PyTorch also provides a JIT compiler called TorchScript, which returns the Script function \cite{pytorch}. With PyTorch JIT Script, we implement parallelized MAS in two versions, while writing to the path tensor is slower in small tensor size.
Different from Triton and Cython implementation, they calculate batch dimensions together as a parallelization.

\begin{algorithm}[t]
  \caption{Parallelized Monotonic Alignment Search}
  \label{alg_pmas}
\begin{algorithmic}
  \State {\bfseries Input:} log-likelihood matrix $q_{1:T,1:S}$, the speech representation length $S$, the text length $T$
  \State {\bfseries Output:} monotonic alignment $A^*$\\
  \State  \textcolor{igreen}{Initialize the first column $q_{2:S, 1} \leftarrow -\infty$, as each text segment should correspond to at least one speech representation.}
  \State  \textcolor{igreen}{$q_{0,1:S} \leftarrow - \infty$ (omitted in Triton)}
  \For{$j=2$ {\bfseries to} $S$}
  \State \textcolor{igreen}{Inner for loop has been replaced by parallel processing}
  \State \textcolor{igreen}{$q_{1:T,j}  \leftarrow \max(q_{0:T-1,j -  1}, q_{1:T,j - 1}) + q_{1:T,j}$}
  \EndFor
  \State Initialize $A^*(S) \leftarrow T$
  \For{$j=S-1$ {\bfseries to} $1$}
  \State $A^*(j) \leftarrow \mathrm{argmax}_{i \in \{A^*(j+1)-1, A^*(j+1)\}}q_{i,j}$
  \EndFor
\end{algorithmic}
\end{algorithm}

Unlike the Triton kernel, during the forward loop, the JIT scripts slice the $j$th column from $q$ as $q_{1:T,j}$, and use \texttt{torch.roll} to shift the index and masking first index to the maximum negative value. 
\texttt{JIT\_v1} is fully computed with pytorch tensor on the GPU, while \texttt{JIT\_v2} calculates the forward loop on the GPU, copy it to the CPU, backward loop on CPU, and sends the result to the GPU.
Both versions share the same forward loop and differ only in the implementation of the backward loop.

Recently, PyTorch has supported static graph compilation for dynamic shapes \cite{pytorch_dynamic}. 
However, applying multiple static graph compilations is challenging for TTS tasks, as both the text length and the speech length vary. Therefore, we did not attempt a compiled implementation in our experiments.

\section{Experiments}
To evaluate the performance of our Super-MAS kernel and JIT scripts, we run benchmarks in 
the identical system. 
The benchmark was run on a system with Intel Xeon Gold 6226R CPU and NVIDIA RTX 3090 GPU.
The implementation requires PyTorch, tested with version \texttt{torch==2.3.0+cu121}, Triton-Lang, tested with version \texttt{triton==2.3.0}, and Cython, tested with version \texttt{Cython==0.29.36}.

During the benchmark, we generate a random log-likelihood tensor of size $[B,T,S]$, where $B$ is the batch size, $T$ is the text length, and $S$ is the speech length.
We fix $B$ to 32, and $S$ to four times $T$ from the static of LibriTTS dataset.
This constraints make $T$ being the only variable during the benchmark, and the time complexity of the algorithm to $O(T^2)$.
The benchmark was conducted using Triton's \texttt{triton.testing.perf\_report}. 
For each test, we warm up the system with 25 steps and then measure the average time over 500 repetitions.
Please refer to \href{https://github.com/supertone-inc/super-monotonic-align}{our GitHub repository} for the detailed setup and benchmark processes.

\section{Results}


Table \ref{tab:result} and Figure \ref{fig:result} illustrate the execution time results obtained from the benchmark. 
The Super-MAS Triton implementation demonstrates a remarkable performance improvement by being at least 19 times faster, and in extremely long cases, up to 72 times faster than the Cython implementation. 
This substantial speedup is achieved through an optimized kernel design that parallelization of the algorithm, efficiently reuses the memory space of the provided log-likelihood matrix, thereby eliminating unnecessary memory allocation and overhead.

In addition to the Triton implementation, the PyTorch JIT implementations also outperform the Cython implementation, particularly when handling large-sized tensors. 
Notably, version v1 of the PyTorch JIT implementation exhibits superior performance because it avoids any inter-device data copying between CPU and GPU memory. 
This design choice minimizes the communication overhead that can otherwise degrade performance when processing large volumes of data.

As shown in Figure \ref{fig:result}, other implementations exhibit fluctuating runtimes and cannot be treated as line on the log-scale plot.
However, Triton implementations forms a straight line in the log-scale graph which is indicating stable time expectations and minimal overhead.

Overall, these results underscore the effectiveness of our implementations in accelerating MAS computations during TTS model training. The significant reduction in execution time not only enhances computational efficiency but also contributes to a considerable decrease in overall training time, which is especially beneficial given that TTS models are typically almost millions of iterations. 

For a rough estimation, consider Glow-TTS, which is trained on LibriTTS \cite{libritts} and uses the same batch size as our benchmark \cite{glowtts}. 
According to Kim \textit{et al.} \cite{glowtts}, the MAS computation in Glow-TTS takes less than 20 ms. Assuming that the average iteration involves cases where $128 < T < 256$, and given that the maximum text length is reported as 190 characters \cite{glowtts}, we can derive an upper bound for the MAS time difference as 20 \text{ms} - 1.617 \text{ms} = 18.383 \text{ms}. 
Since the multi-speaker model was trained for 960k steps, this results in a total time saving of approximately less than 4.4 hours in 12 days, which corresponds to reduce 1.5\% of total training time.
Since this calculation is based on a subset of the LibriTTS dataset containing samples with text lengths shorter than 190 characters, the estimated time savings are reflective of performance on relatively short sequences.
In practice, when our method is applied to longer speech datasets that include extended text sequences, such as LibriSpeech-Long \cite{librilong}, the computational load increases as quadratic scales, leading to significantly greater time savings due to the higher overhead associated with processing longer inputs.








\section{Conclusion}
In this paper, we implemented parallelized MAS in multiple frameworks and versions.
Our Super-MAS Triton kernel shows the best time, achieving at least 19 times faster and up to 72 times performance than the original Cython implementation.
Since MAS is computed at every iteration on-the-fly during TTS model training, which typically involves almost 1 M steps, this kernel can reduce the overall training by at least 1.5 \%.
This result underscore the critical role of GPU-level optimization in modern speech and TTS systems, where even seemingly small computation components can become bottlenecks without careful kernel design.
Our kernel, which is focused solely on optimizing MAS, does not currently incorporate kernel fusion techniques for calculating the log-likelihood or for multiplying the MAS path with the text to generate the aligned text-grid.
Therefore, there remains significant potential for further optimizing the training process of TTS models by leveraging MAS.
We believe this work can be utilized in various applications, including future non-autoregressive TTS models, alignment estimation for automatic speech recognition models, and other scenarios requiring monotonic alignment.

\bibliographystyle{IEEEbib}
\bibliography{mybib}

@inproceedings{glowtts,
 author = {Kim, Jaehyeon and Kim, Sungwon and Kong, Jungil and Yoon, Sungroh},
 booktitle = {NeurIPS},
 pages = {8067--8077},
 title = {{Glow-TTS: A Generative Flow for Text-to-Speech via Monotonic Alignment Search}},
  volume={33},
 year = {2020}
}

@inproceedings{vits,
  author={Jaehyeon Kim and Jungil Kong and Juhee Son},
  title={{Conditional Variational Autoencoder with Adversarial Learning for End-to-End Text-to-Speech}},
  year={2021},
  pages={5530-5540},
  booktitle={ICML},
}

@inproceedings{flashattn,
 author = {Dao, Tri and Fu, Dan and Ermon, Stefano and Rudra, Atri and R\'{e}, Christopher},
 booktitle = {NeurIPS},
 pages = {16344--16359},
 title = {FlashAttention: Fast and Memory-Efficient Exact Attention with IO-Awareness},
 volume = {35},
 year = {2022}
}

@inproceedings{
flashattn2,
title={FlashAttention-2: Faster Attention with Better Parallelism and Work Partitioning},
author={Tri Dao},
booktitle={ICLR},
year={2024},
url={https://openreview.net/forum?id=mZn2Xyh9Ec}
}

@inproceedings{
nansy,
title={{Neural Analysis and Synthesis: Reconstructing Speech from Self-Supervised Representations}},
author={Hyeong-Seok Choi and Juheon Lee and Wansoo Kim and Jie Hwan Lee and Hoon Heo and Kyogu Lee},
booktitle={NeurIPS},
year={2021},
}

@inproceedings{
nansypp,
title={{NANSY}++: Unified Voice Synthesis with Neural Analysis and Synthesis},
author={Hyeong-Seok Choi and Jinhyeok Yang and Juheon Lee and Hyeongju Kim},
booktitle={ICLR},
year={2023},
}

@inproceedings{dualspeech,
  title     = {{DualSpeech: Enhancing Speaker-Fidelity and Text-Intelligibility Through Dual Classifier-Free Guidance}},
  author    = {Jinhyeok Yang and Junhyeok Lee and Hyeong-Seok Choi and Seunghoon Ji and Hyeongju Kim and Juheon Lee},
  year      = {2024},
  booktitle = {Interspeech 2024},
  pages     = {4423--4427},
  doi       = {10.21437/Interspeech.2024-2005},
}

@inproceedings{
pits,
title={{PITS}: Variational Pitch Inference Without Fundamental Frequency for End-to-End Pitch-Controllable {TTS}},
author={Junhyeok Lee and Wonbin Jung and Hyunjae Cho and Jaeyeon Kim and Jaehwan Kim},
booktitle={ICML 2023 Workshop on Structured Probabilistic Inference {\&} Generative Modeling},
year={2023},
}

@inproceedings{triton,
author = {Tillet, Philippe and Kung, H. T. and Cox, David},
title = {Triton: an intermediate language and compiler for tiled neural network computations},
year = {2019},
booktitle = {Proceedings of the 3rd ACM SIGPLAN International Workshop on Machine Learning and Programming Languages},
pages = {10–19},
series = {MAPL 2019}
}

@ARTICLE{cython,
  author={Behnel, Stefan and Bradshaw, Robert and Citro, Craig and Dalcin, Lisandro and Seljebotn, Dag Sverre and Smith, Kurt},
  journal={Computing in Science \& Engineering}, 
  title={Cython: The Best of Both Worlds}, 
  year={2011},
  volume={13},
  number={2},
  pages={31-39},
}

@inproceedings{pytorch,
author = {Paszke, Adam and Gross, Sam and Massa, Francisco and Lerer, Adam and Bradbury, James and Chanan, Gregory and Killeen, Trevor and Lin, Zeming and Gimelshein, Natalia and Antiga, Luca and Desmaison, Alban and K\"{o}pf, Andreas and Yang, Edward and DeVito, Zach and Raison, Martin and Tejani, Alykhan and Chilamkurthy, Sasank and Steiner, Benoit and Fang, Lu and Bai, Junjie and Chintala, Soumith},
title = {PyTorch: an imperative style, high-performance deep learning library},
year = {2019},
booktitle = {NeurIPS},
articleno = {721},
numpages = {12}
}

@manual{nvidia_ampere,
  title        = {NVIDIA AMPERE GA102 GPU Architecture},
  author       = {NVIDIA Corporation\phantom{}},
  year         = 2020,
  note         = {\url{
      https://www.nvidia.com/content/PDF/nvidia-ampere-ga-102-gpu-architecture-whitepaper-v2.pdf}},
}

@manual{nvidia_a100,
  title        = {NVIDIA A100 Tensor Core GPU Architecture},
  author       = {NVIDIA Corporation},
  year         = 2020,
  note         = {\url{
      https://images.nvidia.com/aem-dam/en-zz/Solutions/data-center/nvidia-ampere-architecture-whitepaper.pdf}},
}

@misc{cuda,
  author={NVIDIA and Vingelmann, Péter and Fitzek, Frank H.P.},
  title={CUDA, release: 10.2.89},
  year={2020},
  url={https://developer.nvidia.com/cuda-toolkit},
}

@inproceedings{transformer,
 author = {Vaswani, Ashish and Shazeer, Noam and Parmar, Niki and Uszkoreit, Jakob and Jones, Llion and Gomez, Aidan N and Kaiser, \L ukasz and Polosukhin, Illia},
 booktitle = {NeurIPS},
 pages = {},
 title = {Attention is All you Need},
 volume = {30},
 year = {2017}
}

@inproceedings{libritts,
  author={Heiga Zen and Viet Dang and Rob Clark and Yu Zhang and Ron J. Weiss and Ye Jia and Zhifeng Chen and Yonghui Wu},
  title={{LibriTTS: A Corpus Derived from LibriSpeech for Text-to-Speech}},
  year=2019,
  booktitle={Proc. Interspeech},
  pages={1526--1530},
  doi={10.21437/Interspeech.2019-2441}
}

@misc{pytorch_dynamic,
  author={PyTorch Docs},
  title={Dynamic shapes},
  year={2024},
  url={https://pytorch.org/docs/stable/torch.compiler_dynamic_shapes.html},
}

@article{librilong,
  author       = {Se Jin Park and
                  Julian Salazar and
                  Aren Jansen and
                  Keisuke Kinoshita and
                  Yong Man Ro and
                  R. J. Skerry{-}Ryan},
  title        = {Long-Form Speech Generation with Spoken Language Models},
  journal      = {CoRR},
  volume       = {abs/2412.18603},
  year         = {2024}
}

\end{document}